**Investigating Thermal Boundary Conductance in Monolayer Transition Metal Dichalcogenides: A Systematic Comparison of Advanced Nanoscale Thermometry Techniques**

*Timm Swoboda*[1] | *Jiaqi Yang*[1] | *Arkadiusz P. Gertych*[2] | *Carlos Acosta*[3] | *Onurcan Kaya*[1] | *Daniel Capolat Palomar*[1] | *Severine Gomès*[3] | *Aitor Lopeandia*[1,4] | *Stephan Roche*[1] | *Javier Rodriguez-Viejo*[1, 4] | *Marianna Sledzinska*[1*]

[1]Catalan Institute of Nanoscience and Nanotechnology (ICN2), CSIC and BIST, Barcelona, 08193, Spain | [2]Faculty of Physics, Warsaw University of Technology, 00-662 Warsaw, Poland | [3] CNRS, INSA Lyon, CETHIL, UMR5008, 69621 Villeurbanne, France | [4]Physics department, Autonomous University of Barcelona (UAB) Campus UAB, Bellaterra, Barcelona, Spain

**Correspondence:** Timm Swoboda (timm.swoboda@icn2.cat) | Marianna Sledzinska (marianna.sledzinska@icn2.cat)

**Funding:** M. S, T. S and J. R. V. acknowledge the support from the projects 3D-Bricks (Horizon Europe, GA No. 101099125), PETITE (GA No. PCI2023-143399). A.P.G. acknowledges the financial support within Etiuda Grant funded by the Polish National Science Center (2019/32/T/ST3/00318).

**Graphical Abstract**

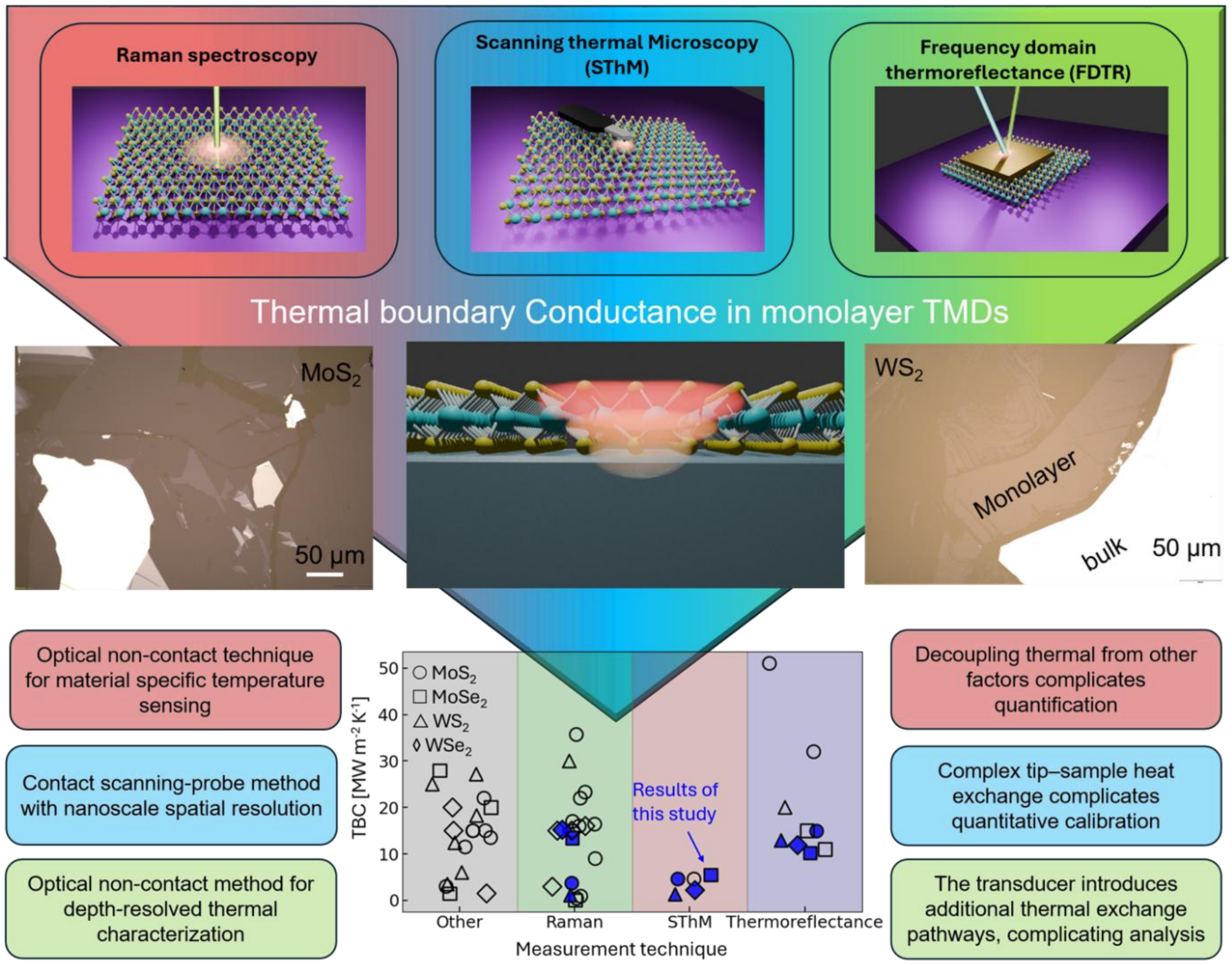

**Abstract**

Thermal boundary conductance (TBC) plays a crucial role in heat dissipation in two-dimensional (2D) devices. However, reported values for 2D material classes such as monolayer transition metal dichalcogenides (TMDs) can vary by more than an order of magnitude across the literature, making it difficult to determine whether these differences come from the materials themselves or from the measurement methods used. To address this issue, this article presents a systematic, cross-technique study of the interfacial heat transport in four TMD monolayers supported on a sapphire substrate, using three of the most representative methods: Raman thermometry, scanning thermal microscopy (SThM) and frequency domain thermoreflectance (FDTR). By conducting all measurements on the same sets of samples, the influence of sample-to-sample variation is minimized allowing for a direct comparison of the three approaches. The results reveal a variation in the measured values depending on both the material and the measurement technique. By directly comparing these results we can elucidate on the origin of the differences and underpin sources of uncertainty in each of the characterisation methods. Taken together, these results provide a direct experimental benchmark for studies of 2D material interfaces and offer practical guidance for the selection and interpretation of thermal metrology.

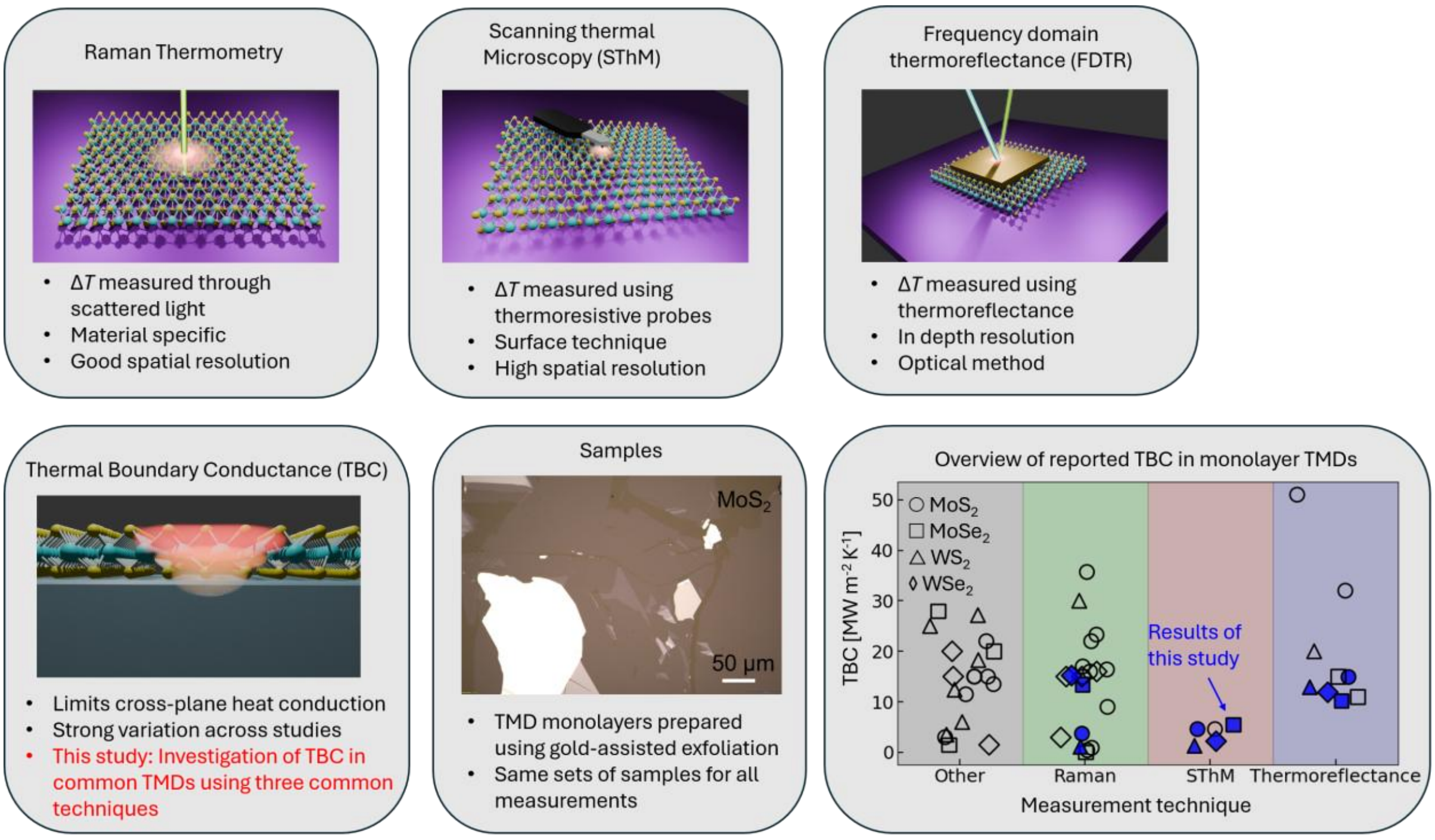

**Figure 1.** Schematical illustration of the used measurement techniques. Overview of reported thermal boundary conductance values between monolayer transition metal dichalcogenides (TMDs) and their substrate ($SiO_2$, sapphire or AlN, blue bulk symbols display results of this study).

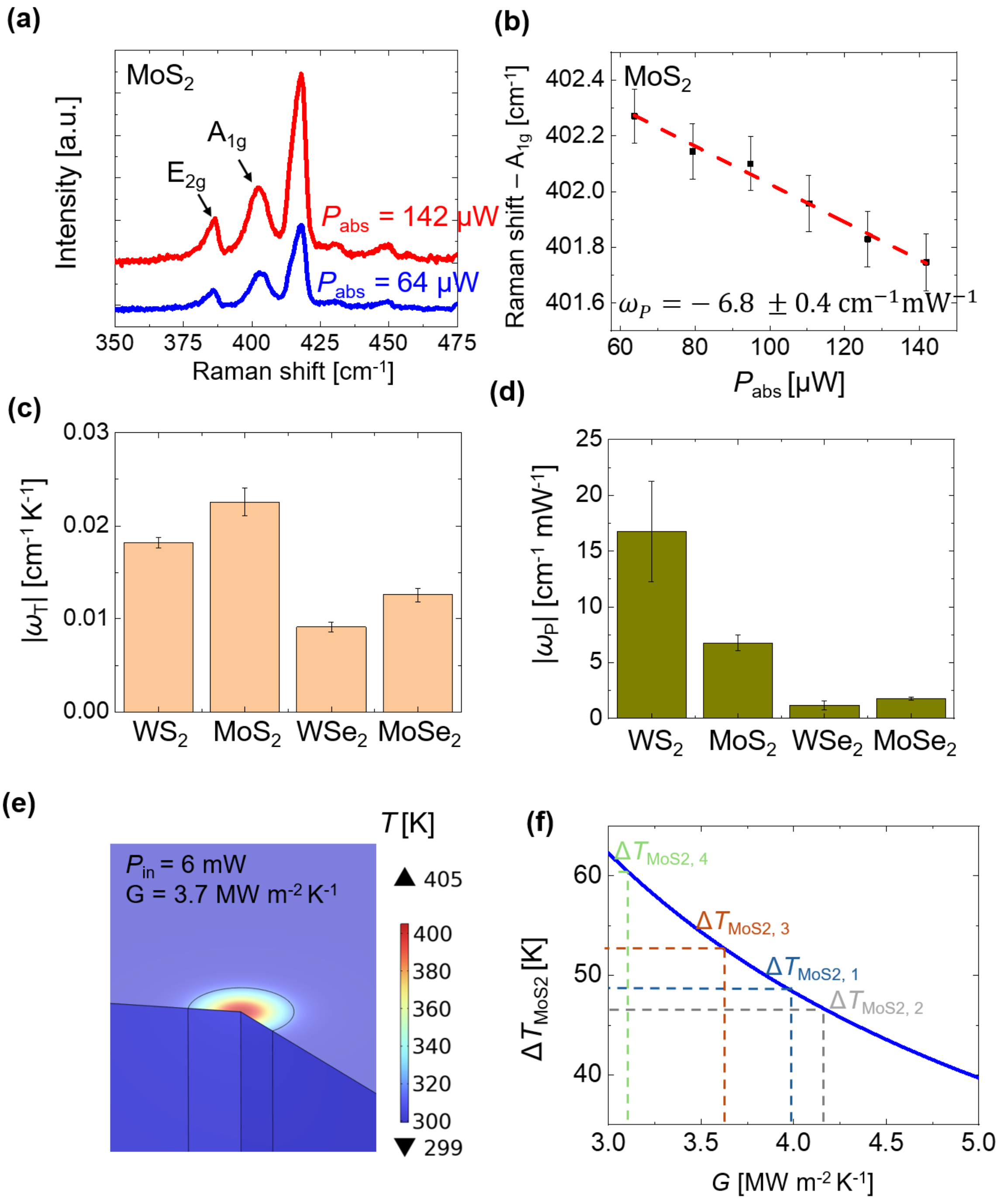

**Figure 2.** Results of Raman thermometry measurements **(a)** Raman spectra of a $MoS_2$ monolayer for two different laser power magnitudes. **(b)** Fitted Raman shift peak position of the $A_{1g}$ peak in $MoS_2$ as a function of the absorbed laser power, **(c)** Bar plot of the temperature calibration magnitude of the Raman shift $\omega_T$ for the four studied monolayers, **(d)** Bar plot of the power dependency magnitude of the Raman shift $\omega_P$ for the four studied materials, **(e)** Finite element method (FEM) simulation showing the temperature $T$ in a heated $MoS_2$ monolayer, **(f)** Average temperature increase $\Delta T_{MoS2}$ of the heated laser spot area as a function of the model parameter G for $MoS_2$, the solid line shows the results of the FEM model, while the experimental values represented by the dashed lines are used for fitting.

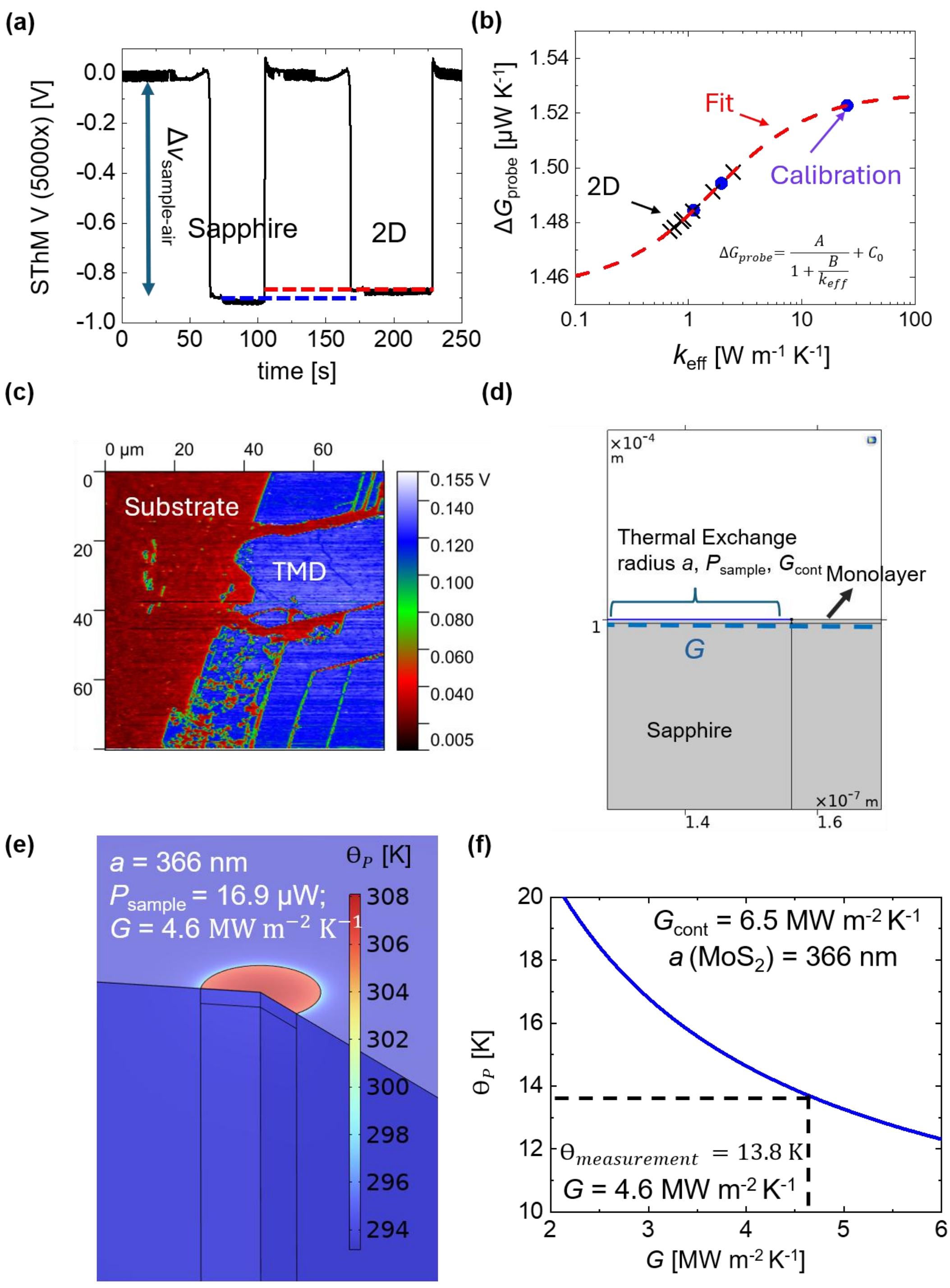

**Figure 3 (a)** Graphical illustration of the scanning thermal microscopy (SThM) logger data as a function of the time showing the signal change when bringing the probe in contact with the surface of the studied material, **(b)** Calibration curve showing the fitted changes in the SThM probe thermal conductance $\Delta G_{probe}$ as a function of the thermal conductivity $k_{eff}$ of the studied samples, Blue dots represent the results of the calibration samples, the dashed line represents the calibration fit (A = 0.071 µW K-1, B = 1.699 W m-1 K-1, A = 1.457 µW K-1) and the black crosses represent the results obtained for the TMDs. **(c)** SThM scan showing the contrast between a monolayer $WS_2$ and the sapphire substrate, **(d)** Geometry of the finite element method (FEM) model used for the calculation of the thermal boundary conductance, **(e)** Temperature distribution in a heated $MoS_2$ monolayer using the thermal exchange area $a$ = 366 nm, **(f)** Average Probe temperature $\Theta_P$ of the heated thermal exchange area as a function of the model parameter *G* the solid line shows the results of the FEM model, while the experimental value represented by the dashed line is used for fitting**.**

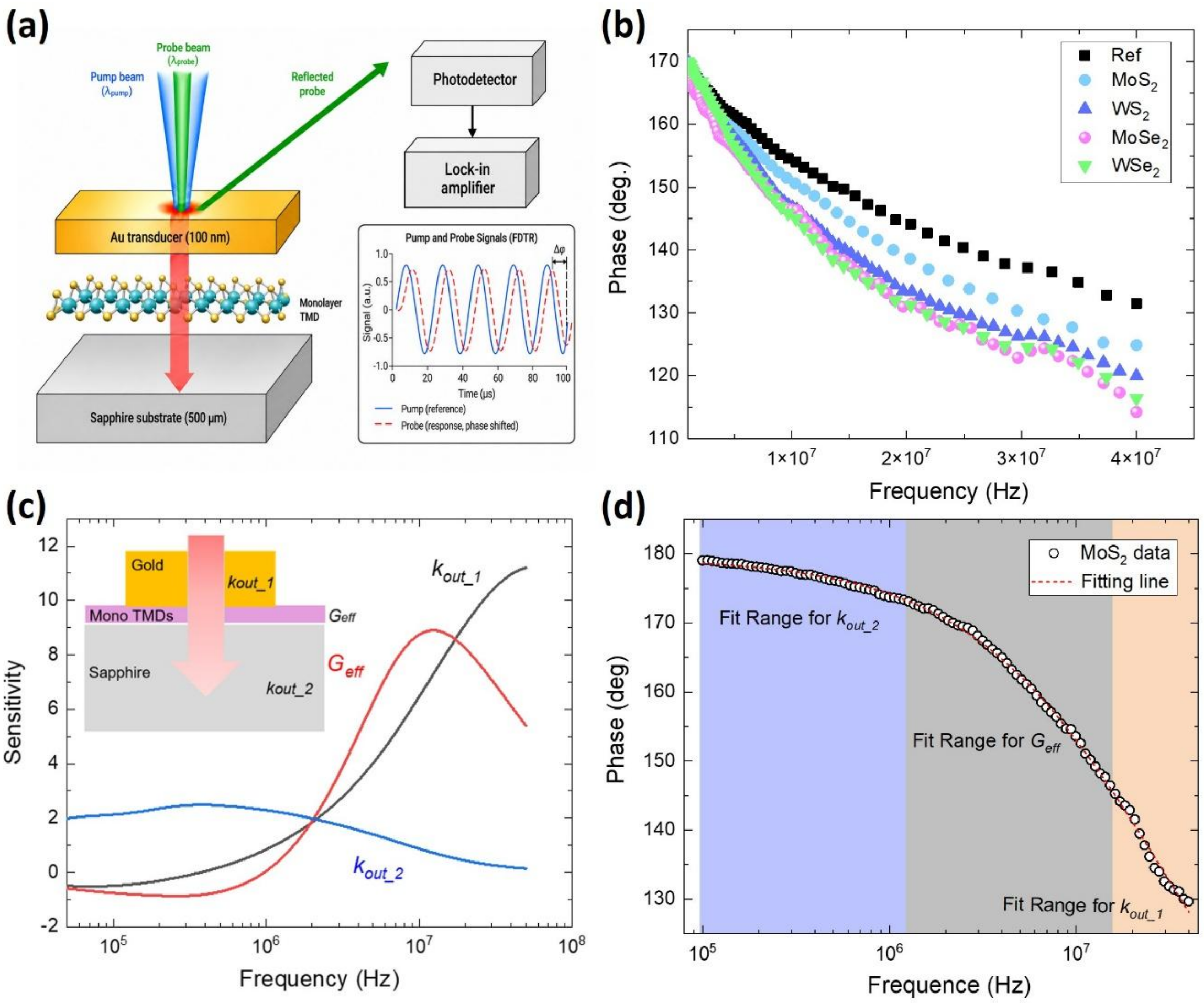

**Figure 4** Frequency domain thermoreflectance (FDTR) measurements and extracted thermal boundary conductance (TBC). **(a)** Schematic of the FDTR measurement configuration for a Au/monolayer TMD/sapphire stack. **(b)** Phase vs. modulation frequency for the Au/monolayer-TMD/sapphire stacks (monolayer $WS_2$, $WSe_2$, $MoS_2$, $MoSe_2$) and the Au/sapphire reference. **(c)** Frequency-dependent sensitivity of FDTR thermal parameters ($k_{out_1}$, $G_{eff}$ and $k_{out_2}$) for an Au-Monolayer TMD-sapphire system. **(d)** Fitting example: extracted effective TBC between Au, Mono MoS2 and Sapphire (symbols: data; line: model).

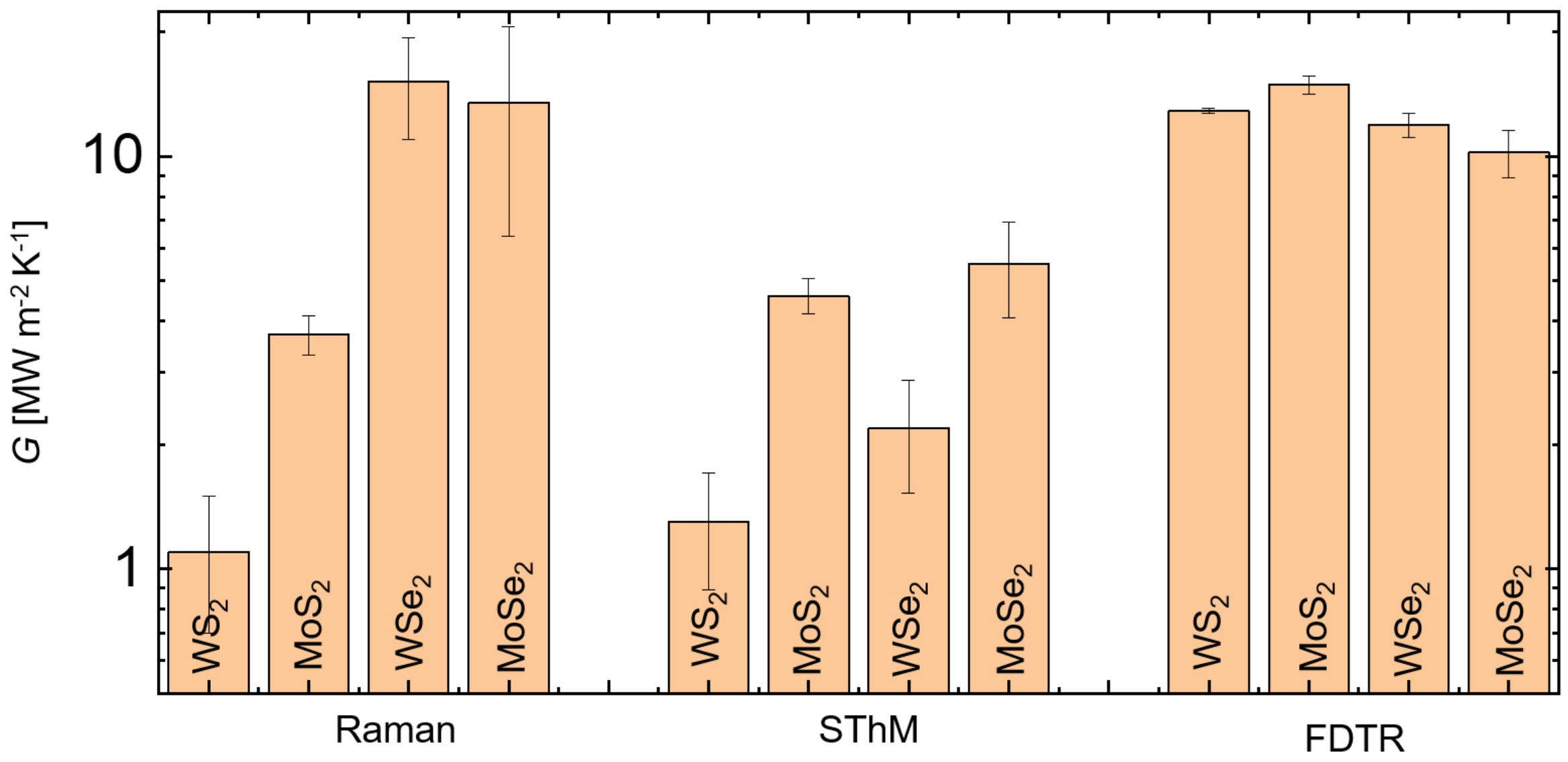

**Figure 5** Overview of measured thermal boundary conductance between the studied monolayers and the sapphire substrate from left to right obtained by Raman thermometry, Scanning thermal microscopy (SThM) and Frequency domain thermoreflectance (FDTR) measurements.

## 1 | Introduction

Over the past decade, research on two-dimensional (2D) transition metal dichalcogenides (TMDs) has increased significantly.[1–3] In particular the characterization of thermal properties becomes increasingly relevant since these materials display great characteristics for modern thermal management in electronic devices.[1,4] As a result, numerous experimental and theoretical studies have focused on characterizing the thermal transport properties in different 2D TMD systems.[5–7] The most common measurement techniques range from optical methods such as Raman thermometry and thermoreflectance, to contact methods such as scanning thermal microscopy (SThM) or thermal bridge.[5,7–9] These methods are well established and have been discussed extensively in literature.[10–12]

Nonetheless, cross comparisons between such studies remain challenging since the reported values vary depending on the material, the sample preparation, the characterization tool etc. [4] As a consequence, thermal conductivity *k* and TBC have been reported over a large range of values.[13] For example, [4] in-plane thermal conductivity values ranging from 30 to 150 W $m^{-1}$ $K^{-1}$ have been reported for monolayer (1L) $MoS_2$ in different studies.[4] For the interfaces of TMD monolayers ($MoSe_2$,[14–17] $WSe_2$, [14,15,18–22] $WS_2$,[14,15,21,23,24], $MoS_2$[5,7,14,15,17,18,23–29]) to their substrate, the range of reported values spans from below 1 MW $m^{-2}$ $K^{-1}$ up to 50 MW $m^{-2}$ $K^{-1}$.[5,14–30] Figure 1 shows previously reported TBC values in 1L- TMDs including the results obtained in this study. Although the TBC varies naturally depending on the sample substrate,[14,30] even when comparing material systems of the same substrate their TBCs have displayed significant variation amongst the studies. The TBC characterization remains a highly relevant topic in different fields of application. On the one hand, a low TBC can be seen as a drawback due to the reduction of energy transport in electronics. On the other hand, a low TBC system can be desired for applications as heat barriers.[7,22]

In this study, we characterize and compare the TBC in four different 1L-TMDs ($MoS_2$, $MoSe_2$, $WS_2$, $WSe_2$) on sapphire using three different characterization tools (Raman thermometry, frequency domain thermoreflectance (FDTR) & SThM). Since each of these techniques have their own challenges and potential sources of errors, all measurements are performed on the same TMD flakes, to reduce sample-related variation and enable in-depth comparison between the techniques. By that means we explore differences in the measurement physics and results between these tools and gain valuable insights in the benefits and challenges of each technique. Measurements were repeated at different times in order to verify that differences are not stemming from time related contaminations of the samples. We observed both differences in the magnitude of the TBC depending on the measurement technique as also a variation in the TBC of the four monolayers. We elaborate on the origins of these differences and explore the key parameters when comparing the techniques.

Here, we first, elaborate on the fabrication and structural characterization of the monolayers. Second, we introduce the three measurement techniques and the experimental methodology. Then, we benchmark the obtained results and discuss potential implications on the materials and the measurement techniques. Finally, the results are compared to molecular dynamics simulations. Figure 1 shows a schematical illustration of the concept of the used measurement approaches.

## 2 | Materials and Methods

### 2.1 | Sample fabrication and characterization

The samples were prepared using a gold-assisted exfoliation process.[31] The bulk TMDs ($MoS_2$, $MoSe_2$, $WS_2$, $WSe_2$) were purchased from a 2D semiconductor company. Next, the gold tape was prepared by evaporating a 100 nm thick gold layer onto a silicone substrate and covering it with a protective PMMA layer. The substrate was then divided into small pieces, and thermal release tape was used to lift off a clean and smooth gold surface (gold tape stack). Immediately afterward, the gold tape was brought into contact with a freshly uncovered bulk surface. Subsequently, the monolayer was transferred onto a sapphire substrate, and the layers of the gold tape stack were removed sequentially. The thermal release tape was removed by placing the sample on a hotplate for 10 minutes. The PMMA layer was removed using an acetone bath followed by mild $O_2$/Ar plasma treatment. The gold layer was removed using a KI/$I_2$ solution (Sigma-Aldrich standard gold etchant), followed by rinsing with $H_2O$. Finally, the sample was placed under high vacuum and annealed at 350 °C for 3 hours.

Based on that we obtain large scale TMD monolayers for all the four studied materials (see supporting section **S1**). The surface quality of the monolayers was evaluated using atomic force microscopy (AFM). Hereby we observed low surface roughness values between 0.2-0.4 nm which is comparable to the values reported in literature (see supporting section **S2**).[32] We confirmed that the samples are monolayers by measuring the characteristic Photoluminescence (PL) and Raman spectra of the studied materials.[33–40] (see supporting section **S3** & **S4**).

### 2.2 | Raman thermometry

Raman spectroscopy is a powerful and commonly used noncontact tool, employed for the identification and characterization of materials.[41] For temperature measurements, the characteristic Raman vibrational modes are typically used as temperature sensor for the characterization of heat transport properties.[25,42] With increasing sample temperature, the vibrational modes exhibit a redshift and an increased full width at half maximum (FWHM).[17,43–45] By calibrating the redshift, temperature can be quantified. Finite element method (FEM) approaches can be employed to study heat transport parameters on base of the measured temperature increase.[5] In supported monolayer TMDs, heat transport is typically dominated by the cross-plane components, allowing for the quantitative evaluation of the TBC considering a suitable substrate.[5] In suspended samples Raman thermometry can be applied to study in-plane heat conduction.[41,46] Raman thermometry has several limitations, including measurement uncertainties such as the light absorbed by the sample or the laser spot size which are crucial for the quantification.[47] In addition, under laser heating, the Raman-inferred temperature may not accurately represent the true lattice temperature, as different phonon populations can remain out of thermal equilibrium.

In the framework of these measurements, Raman thermometry has the advantage of being a non-contact method and the ability to measure the heating in the structure directly. Furthermore, the identification through the Raman spectra or PL and the actual temperature measurements can be executed within the same setup and position. On the flipside the heat transfer in Raman thermometry is complex. Unlike FDTR or SThM, the power source in Raman thermometry is not external but arises from power absorbed within the material.

In our measurements we use a one laser (532 nm) approach for the measurement of the TBC between the monolayer samples and the sapphire substrate. For the temperature calibration we heat and cool the sample with a Linkam temperature stage and measure the relation between the relevant peak positions and the temperature. For the TBC measurements we heat the sample by increasing the laser power and detect changes in the temperature through the shift of the peak position. All Raman measurements are done in vacuum (< $1x10^{-3}$ mbar).

Figure 2 (a) shows an example of a Raman spectrum of a $MoS_2$ monolayer taken as described above, obtained at two different laser power magnitudes. We fit the spectral position of each peak using a Lorentzian fitting function within the software peak-o-mat. The peak at 383 $cm^{-1}$ corresponds to the $E_{2g}$ mode of $MoS_2$ which represents opposite in-plane vibration of the sulphur atoms in respective to the molybdenum atoms. The peak at 405 $cm^{-1}$ corresponds to the $A_{1g}$ mode of $MoS_2$ and stems from out of plane vibration modes of the sulphur atoms.[48] The peak at 418 $cm^{-1}$ corresponds to the sapphire substrate.[49] Figure 2 (b) shows the dependency of the peak position as a function of the absorbed laser power. The absorbed laser power is measured within the Raman setup. In this case we fit the temperature dependent red shift to lower values induced by the increased laser heating with a linear function with the slope $\omega_P$.

For the calibration of the spectral position shift we heat and cool the sample externally using a liquid nitrogen-cooled temperature stage in 15 °C steps going from -60 °C to 75 °C (see supporting section **S5**). At each temperature we measure the Raman spectrum and fit the peak positions. Subsequently, we calculate the slope $\omega_T$ of the Raman shift vs temperature relation. By dividing the slope of power series by the slope of the temperature calibration we extract the temperature increase as a response to the applied power $P_{abs}$, as follows.

$$\Delta T = \frac{\omega_P}{\omega_T} \cdot P_{abs} \tag{1}$$

Figure 2 (c) shows the fitted $\omega_T$ of the four monolayers. Hereby, we observe a higher magnitude of the slope in $MoS_2$ and $WS_2$ vs $MoSe_2$ and $WSe_2$. This aligns with previous studies and is associated with differences in the strain-phonon coupling.[50] In addition we observe consistent values of the $\omega_T$ from the sapphire substrate verifying the uniformity of the measurement conditions (see supporting section S5).

Figure 2 (d) shows a bar plot of the $\omega_P$ of the four studied monolayers which are calculated as changes per absorbed laser power (see supporting section **S6**). This figure shows a higher $\omega_P$ slope for the sulphur-based samples. This trend indicates a lower heating in the Se-based samples. These values are taken as an average from measurements obtained at five different positions in the sample to take potential local variation into account. In all four monolayers the most dominant peaks are used, i.e. $E_{2g}$ mode for $WS_2$ and $WSe_2$ and $A_{1g}$ for $MoSe_2$ and $MoS_2$.

For the quantification of these results, we employ a heat transfer in solids FEM model using the software COMSOL Multiphysics, where we apply a gaussian heat source to replicate the laser heating of the Raman measurements (see supporting section **S7**). The $1/e^2$ laser spot size is measured using the knife edge method. For the quantification of the heat spreading characteristics, we sweep the TBC between the monolayers and the substrate using a thin layer approximation. For all the simulations we use an arbitrary incident laser power of 6 mW. The absorbed power for the heat source of each material is calculated by multiplying this value with the corresponding absorptance of the monolayer. The thermal conductivity of the layers is taken from literature and present a minor impact on the results as the TBC remains the dominating parameter.[41,51–53] For the fitting of the experimental and theoretical values we calculate the average temperature increase of the monolayer in the laser spot radius area. As the model mimics the measurements, this temperature increase should be equal to the experimentally calculated temperature. Figure 2 (f) shows the temperature results of the FEM simulations as a function of the TBC of $MoS_2$. The dashed lines correspond to the experimentally measured temperatures obtained for the same power at different positions of the $MoS_2$ sample. Following this principle, we average the fitted values for various measurement points for all the four monolayers.

### 2.3 | Scanning thermal microscopy (SThM)

SThM arose as a promising tool for the characterization of heat transport with nanoscale resolution.[11] In this technique a thermoresistive probe is connected to a Wheatstone bridge and heated by applying an electrical current.[11] This electrical network consists of two fixed resistances, an adjustable resistance and the resistance of the probe. The SThM signal is measured as the voltage across the bridge. This configuration allows the SThM to either operate as a temperature sensor or as a temperature heater and a sensor simultaneously.

When used as a heater, a higher current is applied to the Wheatstone bridge to enhance the heat exchange with the sample. In this instance when scanning over a feature with a higher thermal conductivity more heat dissipates from the probe to the sample. By that means the probe will cool down leading to a decrease of the SThM signal. To quantify this signal change, the SThM response is typically calibrated using samples with well-known thermal conductivity, enabling the quantification of the heat spreading.[7,11]

SThM is typically operated in contact mode and thus differs from the other chosen optical techniques. It possesses an excellent spatial resolution depending on the tip size enabling a heating source with nanometric size.[11] This becomes especially relevant for the study of small area samples. Additionally, it can be used to generate fast and accurate 2D maps within a couple of minutes, enabling a strong evaluation of small-scale structures. On the other hand, a careful calibration approach for each probe is required to get quantified information on the sample, which complicates the entire process. In addition, the result depends strongly on the probe to sample interaction. Over time the probes become blunt, which negatively affects the accuracy and comparability when operating with the same probe over an extended period.

In this study we use a KNT probe Model Vita-He-Gla-1 from Bruker with a Pd thermoresistive element in active mode to measure the TBC of the four monolayers. The SThM probe consists of the thermoresistive Pd element, gold pads, wires and two NiCr films (current limiters). The calibration process includes the characterization of the electrical properties of the probe and the SThM response calibration for different reference samples.

For the electrical characterization of the probe, we employ the following strategy (see supporting section **S8**): To extract the probe electrical resistance $R_{Pd}$, we measure the external resistances of the probe separately (gold pads, wires and two current limiters) and subtract the sum from the total resistance $R_{global}$. We measure the temperature dependent $R_{global}$ of the probe in a temperature regulated oven with 10 °C steps. From the linear slope of the corrected $R_{Pd}$ vs temperature dependency we calculate the Pd temperature coefficient of resistance (TCR). In a second step we measure the *IV* curves of the probe both in vacuum and in air. During the SThM measurements we use an electrical current of 0.75 mA. Before we contact our sample, we balance our bridge. Hereby we adjust the controllable resistance so that its value is equal to the probe resistance. According to Kirchhoff's law the SThM voltage is 0 V in the balanced configuration.

For our measurements we employ two approaches, i.e., the logger and scan method. For the logger approach, after balancing the bridge we record the signal of the SThM probe in air. Subsequently we approach the probe to the surface. Once the probe reaches the surface the temperature of the probe reduces due to the additional heat spreading through the thermal exchange between probe and sample. Both the signal in air and the sample in contact is averaged, and the difference represents the voltage drop of the sample $\Delta V$. A larger voltage drop indicates a better thermal exchange between probe and sample as the probe cools down more efficiently. The signal changes are weak as the majority of heat is dissipated through air and the cantilever. To compensate for that, the voltage reading signal is amplified by a factor of 5000. For the scan approach we detect the signal drop as described above at the sapphire substrate. Then we scan along the surface and calculate the average difference in the values obtained at the substrate vs the monolayer (see supporting section **S9**). **Figure 3** (a) shows the logger record taken from a measurement on a $MoS_2$ monolayer. Hereby we observe a stronger drop in the sapphire (blue) in comparison to the monolayer signal (red line). This difference stems from the additional resistance of the $MoS_2$/sapphire interface and thus leads to a reduced heat dissipation.

For the calibration of the signal, we measure the voltage drop in reference bulk materials of known thermal conductivity (including the sapphire substrate). From the voltage drop we calculate the resulting changes in the probe conductance (explained in detail in supporting section **S8**).[7] Subsequently, we fit the changes of the thermal probe conductance $\Delta G_{probe}$ as a function of the effective thermal conductivity of the calibration samples $k_{eff}$ as follows:

$$\Delta G_{probe} = \frac{A}{1+\frac{B}{k_{eff}}} + C_0 \tag{3}$$

where $A$, $B$ and $C_0$ are constants (A = 0.071 μW K$^{-1}$, B = 1.699 W m$^{-1}$ K$^{-1}$, A = 1.457 μW K$^{-1}$).

Figure 3 (b) shows the fitted curve (dashed red) along with the results of the calibration samples (blue dots) and the results obtained in contact with the four monolayers (black crosses). Hereby we can see that the sensitivity of the measurements gets weaker at very high and very low thermal conductivity. At higher thermal conductivity external aspects of the measurements such as the probe to surface contact become more dominant and limit further heat dissipation. At lower thermal conductivity parallel heat spreading through air becomes more dominant and the thermal conductivity of the sample less relevant.

Figure 3 (c) shows a SThM scan taken of the $WS_2$ monolayer along the substrate. Hereby we observe a clear contrast between the signal obtained on the substrate vs the monolayer areas, indicating that the probe is sufficiently sensitive to differentiate between both thermal resistances. We repeat these measurements for all the four monolayers and use both, scans as also the point-wise measurements with the logger option of the system. By that means we explore potential variation in the thermal signal throughout the sample and due to the roughness of the materials. From the calibration curve we then calculate the effective thermal conductivity of the 2D/Sapphire system.

For the quantification of the TBC, we use two different COMSOL FEM models in accordance with the approach of Frausto-Avila et. al.[7] In the first model we assume that the heated probe functions as a power source heating one uniform block of a thermal conductivity of $k_{eff}$ in a thermal exchange radius $a$ (see supporting section **S10**). For that purpose, we use the same power magnitude as during our SThM measurements. Then, we sweep $a$ and calculate the average temperature increase of the sample around this area. We then extract the thermal exchange radius for each sample depending on $k_{eff}$ by matching the calculated temperature from the probe with the temperature of the probe during the measurements, which we calculate from the electrical measurements. In the corresponding simulations we included the thermal boundary conductance $G_{cont}$ between the probe and the sample. For that purpose, we first evaluated the results on the SThM scans. From the ramp region between the 2D materials and the substrate we can estimate the thermal exchange radius of the probe.[54] With the given thermal exchange radius we then swept $G_{cont}$ which served as the value in the subsequent calculations. (see supporting section **S11**)

In the second model we replicate the monolayer/sapphire structure of the real samples using the thermal exchange radius and the thermal contact conductance calculated within the first model (see supporting section **S12**). Figure 3 (d) shows the geometry of the second model including the TBC between the monolayer and the substrate *G*. The same power source is chosen to heat the structure. Figure 3 (e) shows an example of the temperature distribution within the model. In this case we use *G* as the sweep parameter and match the experimental results of the temperature as presented in Figure 3 (f).

### 2.4 | Frequency domain thermoreflectance (FDTR)

In recent years, optical pump–probe thermoreflectance (TR) techniques have been increasingly employed to study the thermal properties of materials, such as the thermal conductivity and the thermal conductance of the interface of multilayer thin films.[10,55] FDTR measurements are based on the relationship between a sample's optical reflectance and its surface temperature. The thermoreflectance coefficient $C_{\mathbf{TR}}$ describes a material's change in temperature $\Delta T$ to the resulting change in optical reflectance ($\Delta R$) using the laser wavelength λ ($R_\lambda$), as follows,[10]

$$\frac{\Delta R_\lambda}{R_\lambda} = (\frac{1}{R_\lambda}\frac{\partial R_\lambda}{\partial T})\Delta T = C_{TR}\Delta T \qquad (4)$$

In FDTR, two lasers are used, one for pumping and one for probing. The frequency modulated pump laser heats the sample periodically. The probe laser is reflected from the heated surface, and the resulting change in surface reflectivity is related to the surface temperature modulation according to equation 4. The temperature oscillation detected by the reflected probe beam follows the same frequency as the pump laser with a phase delay that provides information about thermal properties of the investigated samples as schematically illustrated in **Figure 4** (a).

The main advantage of FDTR lies in its ability to control thermal penetration depth by modulating the pump modulation frequency. As a result, data acquired at different frequencies can be used to extract thermal characteristics at varying depths. For a homogeneous diffusive medium, the thermal penetration depth is approximately given by:[10]

$$\mu = \sqrt{\frac{\alpha}{\pi f}} \qquad (5)$$

(heat penetration depth μ, pump frequency *f* and intrinsic thermal diffusivity α)

Typically, FDTR measurements require the deposition of a metallic transducer (such as Al, Au and Ti) to enhance the otherwise weak thermoreflectance signal. For green lasers Au shows a good reflectivity change with temperature and therefore a 100-nm thick gold transducer layer is used for our measurements.[56]

In our study we use a custom-made FDTR set-up, with a 488 nm pump laser connected to a lock-in amplifier (Zurich Instruments) and a 532 nm probe laser (Cobolt). The phase response was measured over a modulation frequency range of approximately 20 kHz to 40 MHz. For all samples, 100 nm thick Au transducer squares with a lateral size of 20 × 20 μm² were directly deposited by e-beam evaporation on both the monolayer TMD regions and the sapphire reference regions. The TBC of the Au/TMD/sapphire stack was then obtained by fitting an analytical thermal model to the measured phase response.

To study the Au/TMD/sapphire stacks, we adopted an effective two-layer FDTR model (see supporting section **S13 & S14**) consisting of the Au transducer and the sapphire substrate, with a single fitting parameter $G_{\text{eff}}$. The effective TBC can be expressed as series combination of two Kapitza resistances plus the cross-plane thermal resistance of the monolayer $\frac{t_{TMDs}}{k_{out-TMDs}}$ as follows,

$$\frac{1}{G_{eff}} = \frac{1}{G_{Au/TMDs}} + \frac{t_{TMDs}}{k_{out-TMDs}} + \frac{1}{G_{TMDs/Sapphire}} \quad (6)$$

with a thickness of the monolayer $t_{\text{TMDs}} \approx 0.6 - 0.8$ nm and its out-of-plane thermal conductivity ($\boldsymbol{k}_{\text{out-TMDs}}$) of around 1–2 W m$^{-1}$ K$^{-1}$.[34,36,57–59] Based on this the cross-plane thermal resistance of the monolayer is approximately $10^{-10}$ K m$^{-2}$ W$^{-1}$.The TBC values of the two interfaces are within the range of $10^6$ - $5 \cdot 10^7$ W m$^{-2}$ K$^{-1}$. Their inverse values corresponding to the thermal resistance are at least two magnitudes larger than the thermal conduction contribution. Therefore, the extracted $G_{\text{eff}}$ is dominated by the two interfacial contributions.

Figure 4 (b) shows the raw FDTR phase data for four monolayer TMD samples and the sapphire substrate reference. At low frequencies, all samples exhibit a similar phase decrease, whereas at higher frequencies, the sapphire reference exhibits the slowest decay, followed by $MoS_2$ and $WS_2$, with $WSe_2$ and $MoSe_2$ showing similar slopes. The observed differences primarily reflect variations in the effective cross-plane thermal response, which is governed by the effective interfacial TBC of the measured stack. To determine the most suitable frequency window for extracting $G_{\text{eff}}$, we first performed a sensitivity analysis of the FDTR phase response to the key thermal parameters, as shown in Figure 4(c). To minimise uncertainty in the fitted $G_{\text{eff}}$, $k_{\text{out_1}}$ and $k_{\text{out_2}}$were constrained using reported reference values for the Au transducer and sapphire substrate, respectively (the full set of fitting parameters, including layer thickness, volumetric heat capacity, cross-plane thermal conductivity, and thermal anisotropy, is provided in Table **S14**). By that means, parameter correlation is reduced and a more reliable extraction of the effective interfacial TBC is enabled. Figure 4 (d) illustrates this fitting strategy using $MoS_2$ as a representative example, where the phase response is divided into frequency regions mainly sensitive to $k_{\text{out_2}}$, $G_{\text{eff}}$, and $k_{\text{out_1}}$. We therefore extracted $G_{\text{eff}}$ from the mid-frequency range, where the phase response is most sensitive to interfacial heat transport in the Au/TMD/sapphire stack.

## 3 | Results and discussion

To date, the TBC of high-quality, large-area monolayers across all four TMDs has not been systematically investigated under identical experimental conditions, either using a single technique or multiple measurement approaches.

Figure 5 shows the results obtained on the samples investigated in this study. Based on that, different observations can be made in correspondence to the materials and measurement techniques, as elaborated in the following. The Raman results present a larger variation in the results from around 1 MW m$^{-2}$ K$^{-1}$ up to 15 MW m$^{-2}$ K$^{-1}$. The lowest results are observed for the sulphide samples with $WS_2$ exhibiting a value of 1.1 MW m$^{-2}$ K$^{-1}$. Higher values are observed for the selenide samples with the highest value of 15.2 MW m$^{-2}$ K$^{-1}$ in $WSe_2$. The results of the SThM measurements showing comparable values of generally lower magnitude across the monolayers ranging between 1.3-5.5 MW m$^{-2}$ K$^{-1}$. The results of the FDTR display a continuously high magnitude in the absolute TBC values in comparison to the results obtained with the other techniques. Taken the magnitude into account the results show smaller variations in which $MoS_2$ demonstrates the highest $G_{\text{eff}}$ of about 14.4 MW m$^{-2}$ K$^{-1}$ of these four materials, and $MoSe_2$ has the lowest value, at 10.8 MW m$^{-2}$ K$^{-1}$. These results show the same trend as the raw data measured by FDTR. The results for $MoSe_2$ and $WSe_2$ obtained by Raman align well with the values obtained with FDTR.

This outcome supports the claim that the measurement technique plays a critical role in the thermal characterization of the 2D materials. In the following we will discuss the details of the different aspects of each technique that lead to observed differences in the results.

### 3.1 | Raman thermometry

The outcome of our measurements for the sulphides falls within the lower limit of the previously reported results for this technique while the results of the selenides lie in the average region compared to reported values (Figure 1). Interestingly, our results suggest clear differences in the heating of the layers showing increased TBC in the Se-based samples. This was verified with a second set of samples where the same trend was observed. This larger variation between the materials was not observed with the other tools, which could be explained in different ways.

A study of Hunter et. al[20] suggested that radiative recombination of electron hole pairs may lead to an overestimation of the thermal energy in Raman studies of monolayer TMDs. Due to the direct bandgap in these materials, excited electrons have a chance to recombine with the holes in the valence band leading to photon emission. The laser energy used for the excitation of these electrons is therefore not used for heating leading to an overestimation of the heating power in steady state Raman models.[20] The recombination efficiency of this process typically correlates with the PL yield. This would explain the higher TBC in $MoSe_2$ and $WSe_2$ as compared to $MoS_2$. Due to their higher PL yield, the effective heating power would be lower and a higher $\omega_P$ would be observed. However, this would not explain the lower TBC in $WS_2$.

The low TBC values in $WS_2$ could stem from the thermal expansion mismatch between the monolayer and the substrate. Monolayer $WSe_2$ exhibits a thermal expansion coefficient of 9.5 ppm which matches well with the one from Sapphire.[60] On the other hand, monolayer $MoS_2$ (6.7 ppm) and $WS_2$ (6.1 ppm) display lower values and therefore create a bigger thermal expansion coefficient mismatch.[61] In Raman thermometry the laser heating is expected to induce a gap layer emerging due to a thermal expansion mismatch. This gap layer is created along the laser beam energy distribution.[62] This would presumably lead to a locally reduced contact which would explain the lower TBC in materials of a higher mismatch. In addition, temperature differences between the 2D material and the laser heating results in stress due to a large thermal expansion mismatch.[63] Stress inside the material can result in a change of the Raman shift peak position which could be mistaken with a shift induced by temperature. As a consequence, the power slope does not depend on the heating only and therefore would be overestimated. In $MoS_2$ for example this can be evaluated by characterising the $A_{1g}$ vs $E_{2g}$ mode dependency as they show different behavior for heating, doping or strain.[64] In our results the $A_{1g}$ vs $E_{2g}$ varies between heating by temperature stage and by laser heating, while in $WSe_2$ for example the dependency is the same. This discrepancy is a further indicator that in $MoS_2$ the Raman shift is not just temperature dependent. The effects of the thermal expansion mismatch would additionally explain why we did not observe such a trend in the SThM and FDTR measurements as the laser heating only is applied in a strongly confined location.

Specifically, for $WS_2$, one source of error could stem from the investigated Raman mode. The chosen mode at 350 $cm^{-1}$ correlates to an overlap of the $E_{2g}$ mode and the 2LA acoustic mode.[62] As these two modes cannot be analysed separately, we fit them as a shared common peak. Even though we observed a continuous shift of the common peak in the temperature calibration an impact of this on the results cannot be excluded. In other studies, the $A_{1g}$ mode of the $WS_2$ spectrum can therefore be used for analysis.[62] However, as mentioned earlier, this peak is overlapping with the sapphire in our case.

Furthermore, general sources of error in the Raman measurements include the measurement of absorptance, the fitting accuracy and the spatial uncertainty. Another source of error could stem from the z-position of the laser spot in reference to the sample surface.[65] In order to check this behaviour, we investigated the Raman shift as a function of the z position around the focus spot (see supporting section **S18**). From that we observed that the minimum shift position remains relatively constant in proximity to the focus spot. Additionally, the position of the ideal z is only incrementally changing for higher power. Based on this observation, we assume that this factor has only a small impact on the uncertainty of the results. Additionally, the studied characteristics depend on the studied Raman modes and fitting parameter. For example, by using FWHM as a temperature sensor we observed slightly lower magnitudes of the TBC using the same Raman spectra (see supporting section **S19**). Other studies already indicated that the heating behaviour can vary depending on the studied Raman mode.[22,66] However it should be noted that from our results we did not observe any clear trend when comparing to the two used $E_{2g}$ and $A_{1g}$ modes in the materials where these peaks were present corresponding to in-plane and out of plane vibration respectively. A study by Rahbar et al.[66] previously suggested that thermal non equilibrium between optical phonons and acoustic phonons in continuous wave Raman spectroscopy could affect the quantification of the TBC. In order to solve this issue, energy transport state-resolved Raman has been reported to be a promising approach.[12] All this taken into account we can conclude that the results of $MoS_2$ and $WS_2$ suffer from significant uncertainties and or strong local modifications of the contact due to the laser excitation which should be considered in Raman thermometry measurements.

### 3.2 | Scanning thermal microscopy

The values are within a similar magnitude as the one obtained by Frausto Avila et al. [7] Apart from this study there is a limited number of results from the SThM on TMD monolayers. The large error bars of the results originate from the steep correlation of the calibration curve, where small variations cause a strong shift in the effective thermal conductivity reading. Further sources of error are time dependent changes of the tip apex, the thermal contact radius and the contact resistance between the probe and the sample.

Our results show that a substrate with a high thermal conductivity, such as sapphire, exhibit a strong contrast due to the large difference of the effective thermal conduction. Hereby we observed weaker variation in the results in comparison to the Raman thermometry results. Nonetheless, we observe higher magnitudes in $MoS_2$ and $MoSe_2$ over $WS_2$ and $WSe_2$, which could stem from the lower weight of the Mo in comparison to W. However, it should be noted that this could also be explained through the relatively large error, stemming from the steep slope of the calibration. Therefore, in these structures SThM can be used very effectively and rapidly for qualitative measurements, while the measurements error are the major bottleneck for an effective quantitative conversion. One way to decrease the error of the SThM measurements and increase the spatial resolution is by performing the measurements in vacuum to eliminate the air contribution of the thermal exchange between probe and sample.[11,67]

In samples of smaller lateral size SThM can be further beneficial due to its high spatial resolution. On the flipside quantitative errors in the measurements stem from the contact between the probe and sample and through the high uncertainty in the calibration process. Studies can be conducted at different and or higher power values in order to decrease the uncertainty of the results, even though this naturally involves an increased time consumption by both the measurements and the calibration.[68] In addition, the calibration process must be conducted in each probe and is therefore time consuming. [11,67]

While the results present comparable magnitudes between the material the overall observed TBC values are lower than the results from FDTR or for the sulphides in Raman thermometry. Uncertainties arise from several sources, among others the steep shape of the calibration curve that is resulting in potential deviations of the thermal exchange radius, the thermal contact between probe and sample and tip apex degradation during the measurements. In addition, the surface roughness can affect the heat transport between the tip and the sample, which becomes particularly relevant for rougher samples.[67]

### 3.3 | Frequency domain thermoreflectance

We first characterised the Au/sapphire reference area to assess the stability and repeatability of the transducer/substrate interface. The extracted Au/sapphire reference TBC is $G = 27.2 \pm 0.9$ MW m$^{-2}$ K$^{-1}$, (see supporting section **S16**). For comparison, reported room-temperature TBC values for the Au/sapphire interface vary considerably across different measurement approaches, ranging from 20.8 to 65 MW $m^{-2}$ $K^{-1}$. [69–71] In 1993, Stoner and Maris first used picosecond optical technology (time-resolved thermoreflectance) to measure the TBC of the gold/sapphire interface conductance to be 45 MW $m^{-2}$ $K^{-1}$. [70] Li et al. obtained 20.8 MW $m^{-2}$ $K^{-1}$ using the 3ω method,[71] and Jeong et al. [69] measured an average value of approximately 65 ± 10 MW $m^{-2}$ $K^{-1}$ using FDTR. This methodological variability may stem from differences in measurement configuration, sample preparation, and thermal model assumptions. Our measured value falls within this reported range, confirming that the Au/sapphire reference provides a reliable baseline for the subsequent FDTR analysis.

After analyzing the Au/sapphire reference, we applied the same FDTR analysis to the Au/TMD/sapphire stacks. The average effective TBC values obtained from repeated FDTR measurements are summarized in Figure 5: 14.5 ± 1.1 MW $m^{-2}$ $K^{-1}$ for $MoS_2$, 12.7 ± 0.3 MW $m^{-2}$ $K^{-1}$ for $WS_2$, 11.5 ± 1.3 MW $m^{-2}$ $K^{-1}$ for $WSe_2$, and 9.8 ± 1.3 MW $m^{-2}$ $K^{-1}$ for $MoSe_2$ (see supporting section **S15** & **S16**). These values are higher than the SThM results. For the selenides, $WSe_2$ and $MoSe_2$, the FDTR values are closer to those obtained from Raman thermometry, whereas the sulfides show larger discrepancies, particularly $WS_2$. These comparisons should be interpreted in view of the modified heat-conduction pathway probed by FDTR, where the measured response includes not only the TMD/sapphire boundary but also the additional Au/TMD boundary introduced by the transducer.

For monolayer $MoS_2$, Wu et al.[57] used FDTR to measure the total thermal conductance of Au-1L $MoS_2$ on sapphire substrate and reported a value of 19.89 MW $m^{-2}$ $K^{-1}$.[72] Brown et al.[16] used time-domain thermoreflectance (TDTR) to get spatial mapping of the TBC of the metal/1L $MoSe_2$/$SiO_2$ interface. The values range between 10-20 MW $m^{-2}$ $K^{-1}$ depending on the different metal transducer.[65–67] While differences in the substrate, transducer chemistry (reactive vs. inert metals), interface cleanliness, and optical/thermophysical inputs can shift absolute values, our results were within the expected range and consistent across repeated measurements, considering the values discusses above.

FDTR requires the deposition of an Au transducer, which introduces an additional Au/TMD interface, thereby altering the measured heat conduction path. Unlike the additional tip–sample contact resistance in SThM, the metal transducer contact in FDTR can actively modify and, in some cases can have an enhancing effect on the thermal transport.[16] This is because the metal/TMD contact is not only a thermal interface but also an electronically and chemically modified interface. Previous studies have shown that metal contacts can affect the binding energy, charge redistribution, and doping which can result in carrier injection from the metal to the 2D material.[16,73] Velický et al.[74] investigated the interaction between $MoS_2$ and Au and reported a reduction of the binding energies of 0.4 eV in $MoS_2$ monolayer in comparison to bulk $MoS_2$ using X-ray photoelectron spectroscopy. They suggested that this change could stem from significant charge transfer from the Au to the $MoS_2$ leading to an increased electron density at the interface. These additional charge carriers may provide an extra channel for energy transfer, contributing to the larger apparent TBC values observed in metal/TMD/substrate stacks. In addition, the evaporation of the transducer metal could result in a strain in the 2D materials that could affect the thermal properties. Schauble et al.[73] studied the strain in monolayer $MoS_2$ evaporated with different thin metals by means of x-ray diffraction (XRD) measurements. On the contrary to previous reports the authors did not observe any strain in the metal covered $MoS_2$. In addition, their results indicate that strain still does not appear for increasing thickness of the transducer layer.

Another possible factor contributing to the increased TBC is the diffusion of metal atoms into the monolayer.[75] To investigate this, we performed cross-sectional transmission electron microscopy (TEM) imaging in exfoliated $MoS_2$ structures with and without the evaporated Au transducer. In this case no visible diffusion was observed (see supporting section **S17**). Nonetheless it should be noted that the investigated cross sections only represent a small section and even a minor degree of diffusion could influence the measured conductance significantly. Therefore, one mitigation strategy could be the use of transferred transducers instead of evaporated ones as they are expected to exhibit a lower diffusion.[75]

Overall, FDTR is highly sensitive to changes in the effective cross-plane thermal response of layered structures, including 2D material/substrate stacks, but its performance is limited by the influence of the transducer. The Au layer introduces an additional interface and can also modify the local electronic and thermal coupling at the TMD surface. In other studies, particularly those focused on device-relevant structures where a metal electrode is an integral part of the stack, this displays an advantage as the impact of the metal on the heat spreading is already considered. FDTR is further a valid tool for qualitative comparisons of the results as the signal fitting is very sensitive to changes, which could be interesting for studies on few layer structures. Previously, thermoreflectance methods demonstrated to be very effective to characterize differences in few layer 2D materials depending on the number of layers.[24] Sources of error mostly stem from the characterization of the laser spot size, thermophysical input parameters, and fitting procedure.

From a materials perspective, the comparison among $MoS_2$, $WS_2$, $WSe_2$, and $MoSe_2$ should be interpreted with caution. Although the FDTR data show slightly higher effective TBC values for the sulfides than for the selenides, previous studies do not establish a universal ranking for all four monolayer TMDs.

### 3.4 | Materials

From a material standpoint, in the literature there is no clear trend which of the four monolayers should present higher a TBC. For example in a theoretical study the lowest TBC is predicted in $WS_2$ when comparing all the four materials with an amorphous $SiO_2$ substrate.[15] On the other hand a study comparing different substrates shows that $WS_2$ presents the highest TBC when considering a sapphire substrate.[14] This difference indicates that the measured TBC depends not only on the intrinsic phonon properties of the monolayer, but also on the substrate, interface bonding, local contact quality, and the degree of phonon density of states (PDOS) matching across the interface, which will be further discussed through the theoretical calculations below. In general, one could expect a higher thermal conductivity for lighter materials.[76] However our results do not show clear and or consistent trends with the weight of the materials.

When compared with previous reports, our measured TBC values are within the lower range for all the measurement techniques. One possible reason for this could be related to the sample fabrication process. Defects formed during exfoliation or transfer, including folds, wrinkles, residues, and local gaps, may reduce the real contact area at the interface and lead to a lower measured TBC. A study by Yuan et al.[77] suggested that folding type defects originated from the exfoliation process could decrease the TBC in 2D materials. The substrate also plays a significant role. As a cross comparison we measured the TBC of $MoS_2$ monolayers on amorphous $SiO_2$ using Raman thermometry and obtained higher values of TBC of 23.7 MW $m^{-2}$ $K^{-1}$ (see supporting section **S17**). These results emphasise that, when comparing TBC values obtained from different studies, careful consideration must be given to the complete stack configuration, fabrication route, and measurement technique.

### 3.5 | Molecular dynamics

In addition, molecular dynamics simulations were performed to study the TBC between the monolayer TMDs and the sapphire substrate (see supporting section **S20**).[78–81] The resulting TBC values are 31.43 $\pm$ 2.42 MW $m^{-2}$ $K^{-1}$ for $MoS_2$, 27.15 $\pm$ 2.61 MW $m^{-2}$ $K^{-1}$ for $WS_2$, 16.56 $\pm$ 1.98 MW $m^{-2}$ $K^{-1}$ for $MoSe_2$ and 13.34 $\pm$ 2.01 MW $m^{-2}$ $K^{-1}$ for $WSe_2$. The variation in TBC across the four monolayers is mainly governed by the spectral overlap between the TMDC and substrate vibrational modes and the equilibrium spacing between layers.[14,27,82] The commensurate supercell construction imposes a residual strain of 0.3-0.5% on the TMDC monolayers, which shift phonon frequencies slightly, however its impact to the trends we observe is small. The computed values overestimate the experimental results from all three techniques, as the simulations model a perfect crystalline interface free of the voids, contamination and structural defects present in exfoliated and transferred samples.[13,14]

Heat transfer across the interface proceeds through the scattering of out-of-plane flexural acoustic phonons into the vibrational modes of the substrate.[14] The scattering rate is proportional to the substrate density of states and inversely proportional to the square of the phonon frequency, meaning the lowest-frequency ZA modes dominate the total conductance. The efficiency of this transmission depends on the spectral overlap between the low-frequency vibrational density of states of the two-dimensional layer and the substrate.[30,82,83] The phonon density of states extracted from the velocity autocorrelation function shows that $MoS_2$ and $WS_2$ exhibit stronger spectral overlap with sapphire in the 0–8 THz range than the selenides. The heavier mass of Se atoms shifts their vibrational frequencies downward and increases the mismatch with the substrate, reducing the rate of phonon scattering into sapphire.

The physical separation across the van der Waals gap also controls the interaction strength between the materials. The interlayer spacing from the simulations is 3.06 Å for $MoS_2$, 3.11 Å for $WS_2$, 3.34 Å for $MoSe_2$ and 3.48 Å for $WSe_2$. The interfacial coupling stiffness, given by the second derivative of the Lennard-Jones potential at the equilibrium separation, decays steeply with distance.[14] The closer proximity of the sulphides to the sapphire surface therefore produces substantially stronger cross-plane coupling and directly enhances the thermal boundary conductance. Because the same heavier chalcogen that shifts the phonon spectrum also increases the atomic radius and widens the gap, both mechanisms reduce the TBC of the selenides relative to the sulphides.

The calculated values overestimate experimental measurements significantly. Molecular dynamics models simulate a pristine and fully crystalline interface without defects. Real experimental environments introduce physical imperfections such as microscopic voids, polymer residues from transfer processes and substrate surface roughness. These defects scatter phonons diffusively and reduce the effective thermal contact area.

### 3.6 | Outlook on machine learning models

With the advancement of machine learning (ML) models, effective tools for data analysis are currently on an upward trend. In recent years, there have been an increasing number of studies on the application of ML models in Ramana spectroscopy data analysis.[84] Such models could significantly accelerate the speed of data analysis and reduce the associated errors. ML models have, for example, been employed for the classification of material properties or for rapid Raman analysis with high precision.[85,86] Nonetheless, ML studies specifically related to the effect of temperate on the Raman spectra still remain scarce and are yet to be conducted.

One specific example of how ML models could be used to improve the analysis reported here was reported by Dehbashi et al.[87] In this study the, authors developed a ML model for the characterization of thermal conductivity based on SThM measurements. They acquired a large dataset of measurements on different materials with varying thermal conductivity and surface topography. By these means they were able to predict the thermal conductivity across a broader set of samples with high reliability, which could be particularly beneficial for reducing errors related to the surface roughness. In another study by Xiang et al.[88] an ML model was developed to characterize the depth dependent thermal conductivity based on the frequency domain signal in thermoreflectance measurements.

On the one hand, the high potential of ML models is therefore particular promising for enhancing our capabilities in the data analysis of the techniques and materials reported here. On the other hand, further research is required to improve the robustness and generalizability, particularly in the field of temperature related measurements.

## 4 | Conclusions

In conclusion, our results demonstrate that the measured TBC of monolayer TMDs strongly depends on the measurement technique and the specific heat conduction pathway being probed. By applying Raman thermometry, SThM and FDTR to the same set of large area monolayer $MoS_2$, $WS_2$, $MoSe_2$ and $WSe_2$ samples on sapphire, we found systematic differences in both the absolute TBC values and the material dependent trends. FDTR exhibited comparatively high TBC (> 10 MW $m^{-2}$ $K^{-1}$) with small variations between the different materials. SThM resulted in intermediate values (1.3-5.5 MW $m^{-2}$ $K^{-1}$) with small variation between different materials and Raman thermometry produced a large range of values (from around 1 MW $m^{-2}$ $K^{-1}$ for $WS_2$ to above 10 MW $m^{-2}$ $K^{-1}$ for $WSe_2$ and $MoSe_2$) accompanied by larger material dependent spread. These differences highlight the technique specific sensitivities and challenges.